\documentclass[pdflatex,sn-apa]{sn-jnl}

\usepackage{graphicx}%
\usepackage{multirow}%
\usepackage{amsmath,amssymb,amsfonts}%
\usepackage{amsthm}%
\usepackage{mathrsfs}%
\usepackage{fix-cm}
\usepackage[title]{appendix}%
\usepackage{xcolor}%
\usepackage{textcomp}%
\usepackage{manyfoot}%
\usepackage{booktabs}%
\usepackage{algorithm}%
\usepackage{algorithmicx}%
\usepackage{algpseudocode}%
\usepackage{listings}%

\begin{document} 

\title{\centering Intermingled open and closed magnetic field lines\\near the radial origin\\of the heliospheric current sheet}

\author*[1,2]{\fnm{F. S.} \sur{Mozer}}\email{forrest.mozer@gmail.com}
\author[3]{\fnm{A.} \sur{Voshchepynets}}\email{woschep@gmail.com}
\author[1]{\fnm{O. V.} \sur{Agapitov}}\email{agapitov@berkeley.edu}
\author[1]{\fnm{K.-E.} \sur{Choi}}\email{kechoi@berkeley.edu}
\author[1]{\fnm{L.} \sur{Colomban}}\email{lucas.colomban@berkeley.edu}
\author[4]{\fnm{R.} \sur{Sydora}}\email{rsydora@ualberta.ca}

\affil*[1]{\orgdiv{Space Sciences Laboratory}, \orgname{University of California}, \city{Berkeley}, \postcode{94720}, \country{USA}}

\affil[2]{\orgdiv{Physics Department}, \orgname{University of California}, \city{Berkeley}, \postcode{94720}, \country{USA}}

\affil[3]{\orgdiv{Department of System Analysis and Optimization Theory}, \orgname{Uzhhorod National University}, \orgaddress{ \city{Uzhhorod}, \country{Ukraine}}}

\affil[4]{\orgdiv{Physics Department}, \orgname{University of Alberta}, \orgaddress{\city{Edmonton} \postcode{T6G 2E1}, \state{Alberta}, \country{Canada}}}

%

\abstract{
\textbf{Purpose:} To better understand the heliospheric current sheet (HCS) by {describing} the magnetic field geometry{, the plasma, and the} waves in the region near the Sun where the HCS is formed.

\textbf{Methods:} An HCS crossing
having apparent open and closed magnetic field lines
was found and its fields and plasmas were analyzed.

\textbf{Results:} In one hour on 30 March 2024,  the Parker Solar Probe crossed the HCS at 13 solar radii while encountering two distinctly different regions at sharp boundaries on several occasions.  The two regions had very different plasma densities, electric-field spectra, and magnetic-field geometries. In one region the strahl flowed only along the direction from the Sun and there were relatively few particles at pitch angles near 90 degrees, while in the other region the strahl flowed both toward and away from the Sun and there were relatively many particles at pitch angles near 90 degrees.  These different properties are interpreted as being due to the spacecraft crossing into the heliospheric current sheet on long open magnetic field lines in the case of unidirectional strahl flow and the spacecraft moving on to coronal loops having much shorter closed magnetic field lines in the case with bidirectional strahl flow.  The two regions intermingled on time scales less than 100 milliseconds to create a complex magnetic field geometry. Broad-band waves were observed in the open field-line regions while waves observed in both regions were electrostatic harmonics. The harmonic frequencies correlated with the proton plasma frequency, $f_{pp}$, with the lowest frequency at $\sim0.1f_{pp}$. This result, plus the field-aligned electric field waves and plasma density fluctuations, requires that the observed electrostatic mode and associated harmonics were ion acoustic waves. That coronal loops with bidirectional strahl flows were observed as far as 13 solar radii from the Sun during the HCS crossing was not unusual because such loops were observed numerous times during the 12 hours that the spacecraft was in the vicinity of 13 solar radii on this orbit.
}



\maketitle

\section{Introduction}

In the absence of currents in the plasma, the solar magnetic field near the Sun is dipole-like (with an impact from higher-order components) with magnetic field lines that are traceable from/to the solar photosphere and called closed coronal loops. Further from the Sun, magnetic field lines are elongated in the radial direction, leading to the formation of the heliospheric current sheet, which is a mostly longitudinal current that follows the Parker spiral and that separates radial inward pointing from outward pointing magnetic field lines. These field lines form a tail-like structure of elongated field lines (called open field lines, although they must eventually close) \citep{WilcoxNess1965}.  Near the Sun, both open and closed field lines coexist. Processes such as scattering, heating, and acceleration vary with field line length and plasma conditions, shaping the energy and dynamics of each region.

The radial component of the magnetic field measured on the Parker Solar Probe (PSP) \citep{FoxSSRv_2016} can reverse sign either because the spacecraft moves on closed magnetic field lines from one side of the coronal loop to the other or because it {is} 
on longer open magnetic field lines and it moves across the current sheet separating the regions of different polarity \citep{Szabo_2020,Lavraud_2020,Phan_2021GL}. A search {of six orbits was} undertaken to find an example that displays both {open and closed} geometries in a local region{, which is highly unusual}. The FIELDS instrument on PSP measured the electric and magnetic field waveforms at frequencies below about one kHz and collected short bursts of data at frequencies of more than one MHz during an event of interest \citep{BaleSSRv2016}. The electron and proton data in this paper were produced by the SWEAP {and SPAN-I} instruments on the same spacecraft \citep{Kasper_SSRv2016,Whittlesey2020,Livi_2022, Verniero_2022}. All data in this paper are presented in the spacecraft reference frame whose X and Y directions are perpendicular to the Sun-satellite line.
\section{The Data}
    \begin{figure}
   \centering
   \includegraphics[width=9cm]{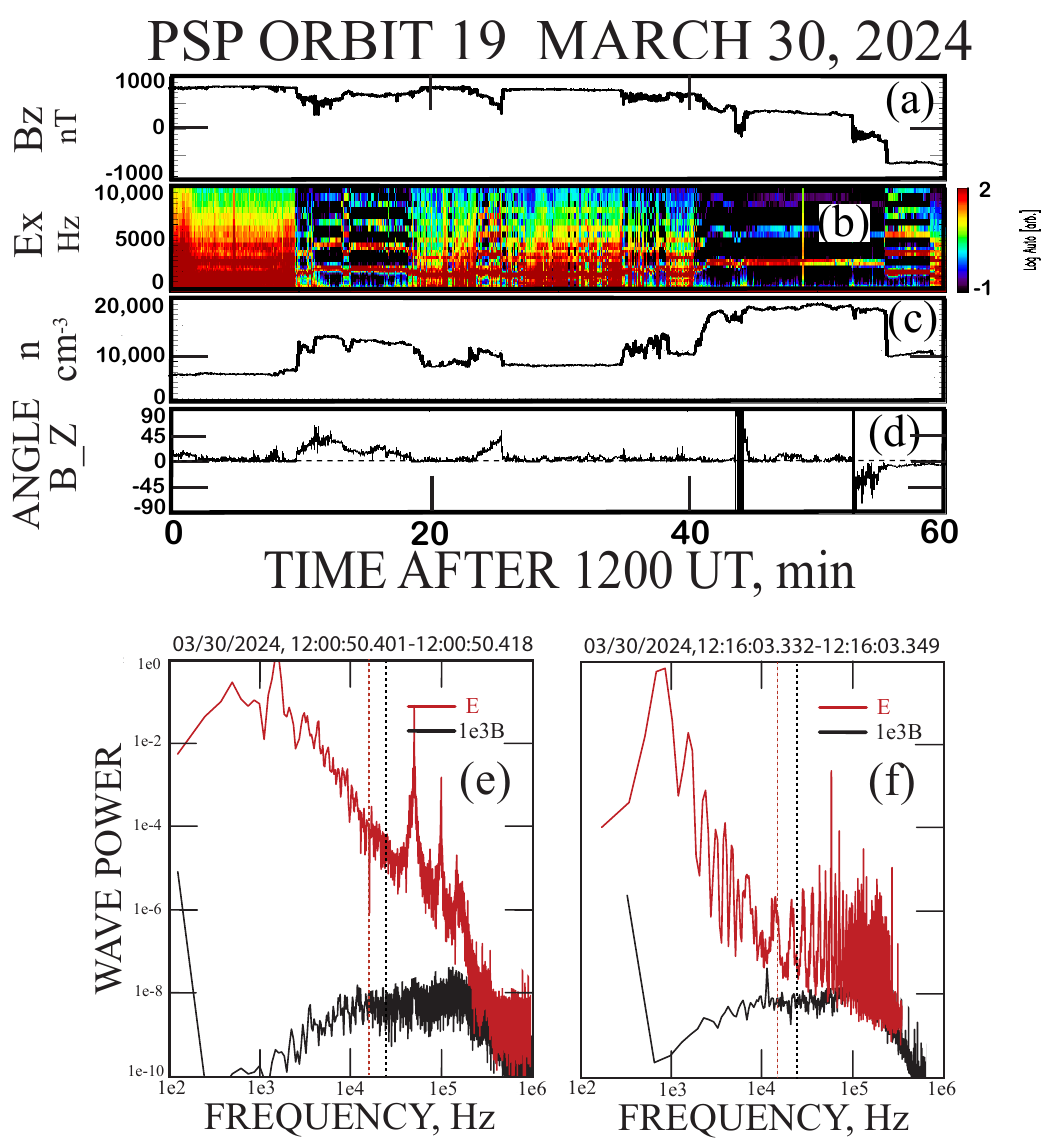}
      \caption{ A one-hour interval during which the radial component of the solar wind magnetic field changed sign (a){, as did the strahl in figure 2a, indicating a crossing of the heliospheric current sheet.} The electric field spectrum in (b) showed two regimes of the observed emissions inside and outside the {heliospheric plasma sheet (HPS)}, (c) - the plasma density changed rapidly from low density outside the {HPS} to the high-density values inside (up to $2\cdot10^4$ cm$^{-3}$), and (d) - the angle of magnetic field deflection from the radial direction varied synchronously with the changes of the plasma density and electric field spectrum, {suggesting a change of the magnetic field geometry}  (e) - spectra during the time interval having a broadband electric field and relatively low plasma density while (f) provides spectra obtained in the region of large density and missing broadband electric field.  The spectra in the latter region are much more complex although they contain less power at frequencies below about 1 kHz.  The absence of magnetic field waves (the black curves in (e) and (f) indicate that all the observed waves are electrostatic.  The black dashed vertical line in both figures gives the electron cyclotron frequency while the dashed red line gives the proton plasma frequency.
              }
         \label{Fig1}
   \end{figure}

  A magnetic field reversal occurred on Parker Solar Probe orbit 19 when the spacecraft was 13 solar radii from the center of the Sun, as illustrated by the radial ($Z$-component) magnetic field change during the one-hour interval given in Figure \ref{Fig1}a. Figure \ref{Fig1}b shows that the electric field spectra consisted of several lines plus a broadband signal that disappeared twice during the crossing (the electron cyclotron frequency was about 10,000 Hz at this time). Figures \ref{Fig1}e and \ref{Fig1}f present the electric and magnetic field spectra obtained to 1 MHz from 17 millisecond bursts that occurred during each of these intervals. The spectrum of Figure 1e was obtained in the region containing the broadband waves while that of Figure \ref{Fig1}f was obtained in the absence of such waves. In addition to the larger power at frequencies below a few kHz in Figure \ref{Fig1}e as compared to Figure 1f, there are much more complex higher frequency waves in the latter case, in part due to the harmonics. In both cases, no magnetic field signature was observed below about 100 kHz. This may be due to either the magnetic field strength being below the threshold of detection or the high-frequency waves being electrostatic. In these figures, the black vertical dashed lines denote the electron cyclotron frequency and the red dashed lines give the proton plasma frequency. It is noted that some of the waves are above and some below the electron cyclotron frequency, as has been previously reported \citep{Mozer_2020,Graham_2021,Piza_2021,Mozer_2021,Tigik_2022,Shi_2022,Malaspina_2022,Ergun_2024}. 

As seen in the plasma density plot of Figure 1c, {(which was verified by the quasi-thermal noise measurements)} the dropouts in the broadband electric field occurred when the plasma density became large, as large as 20,000 cm$^{-3}$. The coordinated sharp boundaries in the electric field spectra and the plasma density suggest the possibility that the spacecraft passed from one region to another at these boundary times. If such boundary crossings meant the passage from open to closed magnetic field lines, the magnetic field should change from near-radial on the open field lines to inclined with respect to the radial direction on the closed field lines. Figure \ref{Fig1}d, a plot of the angle between the magnetic field and the radial direction, shows that the magnetic field acquired a closed-field-line-like geometry when the density increased and the broadband electric field signal disappeared. 

The interpretation that the spacecraft passed between closed and open magnetic field lines may be further tested by considering the flow direction of the strahl, which is a beam of energetic electrons that is formed in the collisional photosphere and that propagates from the Sun \citep{Feldman_1978, Hoeksema_1983, Scudder_1979}.  It is created at both ends of a closed magnetic field line that is anchored in the solar atmosphere.   Anywhere on such closed field lines, called coronal loops, it propagates in both directions.  On open field lines, the strahl signature is a single outgoing beam because the return beam, from the other end of the distant source, travels such a great distance that the strahl signature is lost due to Coulomb collisions, wave interactions, turbulence, etc.  \citep{Halekas_2022}.   Thus, the signature of the strahl is an unambiguous determinant of whether a field line is open or closed. Figure 2a gives the pitch angle distribution of 314 eV electrons, which are a major component of the strahl. It shows that, at times of large plasma density and missing broadband electric field spectra (12:10-12:20 and 12:42-12:55), the near-zero and near-180-degree pitch angle electron fluxes were both large, indicating that the strahl was propagating in both directions along closed magnetic field lines \citep{Lavraud_2020,Szabo_2020,Phan_2021,Desai_2022}. The increases of the electron flux near 90-degree pitch angles in Figure 2a at such times further supports the conclusion that the spacecraft was on closed magnetic field lines at these times because this is the only way to produce trapped particles.  Figure 2b shows the changes of the proton spectra at such times.

\begin{figure}
   \centering
   \includegraphics[width=9cm]{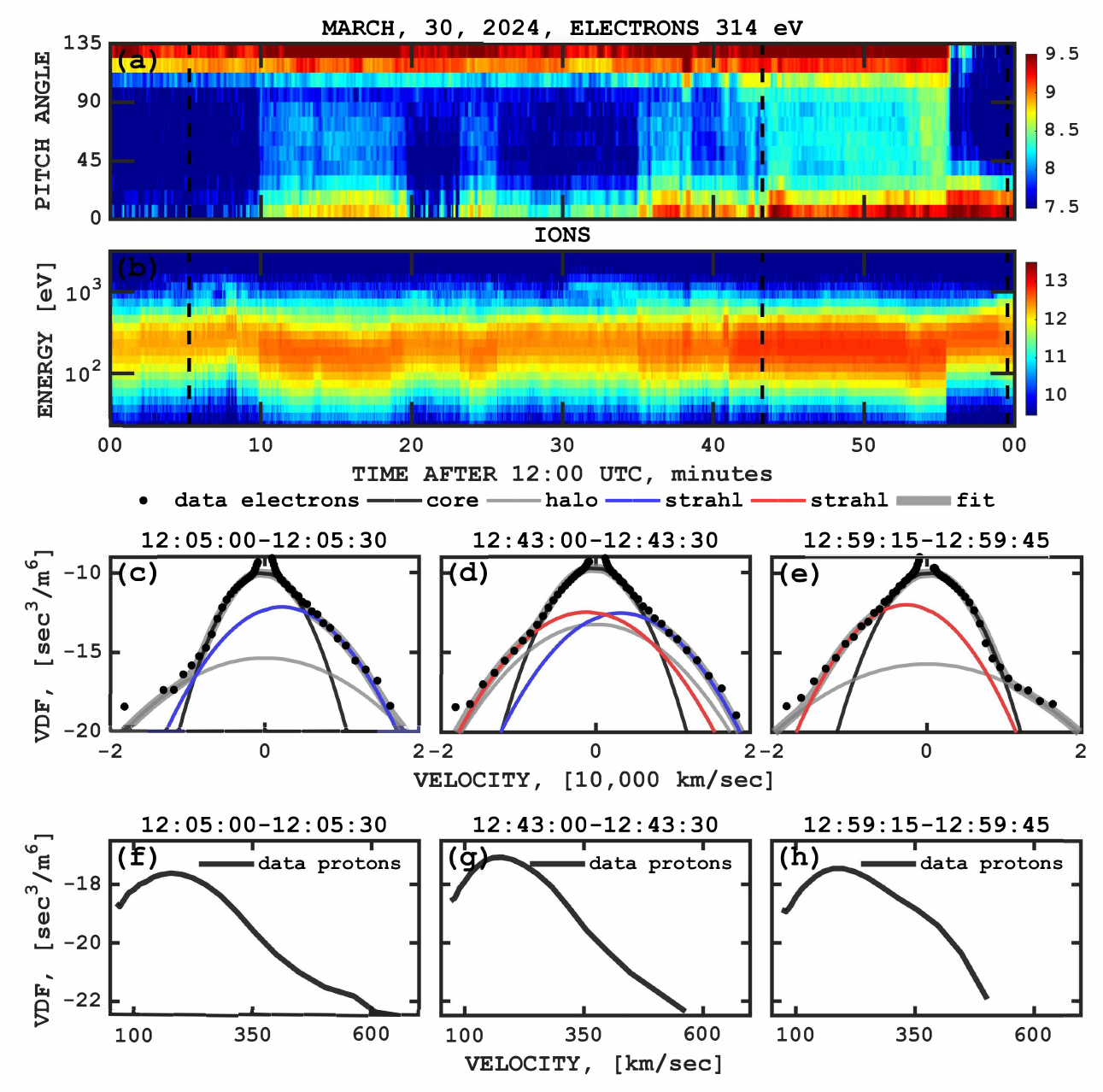}
      \caption{ (a) gives the 314-eV electron pitch angle distribution during the hour of interest, which shows the {counterstreaming} flow of the strahl at 12:10-12:20 and 12:42-12:55, indicating that the spacecraft was on closed magnetic field lines during these times. The enhanced fluxes near 90-degree pitch angles during these times support this interpretation. (b) displays the proton energy distribution while (c), (d), and (e) display one-minute velocity distributions broken-down to display the core, halo, and two strahl distributions at the times of the vertical dashed lines in (a) or (b).  They indicate, respectively, strahl in the minus direction, strahl in both directions, and strahl in the plus direction.  (f), (g), and (h), provide the proton distributions at these same time intervals.
              }
         \label{Fig2}
   \end{figure}

Figures 2c, 2d, and 2e give 30-second averaged electron velocity distribution functions at the times of the vertical dashed lines in Figure 2a or 2b. In each figure, the four dashed curves give the core, halo, and two strahl distributions. They indicate, respectively, strahl in the minus direction at 12:05 on open field lines, strahl in both directions at 12:43 on closed field lines, and strahl in the plus direction at 12:59 on open field lines having a reversed magnetic field. Figures 2f, 2g, and 2h, provide the proton distributions at these same times. The proton density and temperature estimated from the distributions show that the protons had a higher density on the closed field lines (15.9$\cdot$10$^{3}$ cm$^{-3}$ for 2g compared to 5.6 and 9.0$\cdot$10$^{3}$ cm$^{-3}$ for Figures \ref{Fig2} f and \ref{Fig2} h) and a lower temperature (24.2 eV in comparison to 28.2 and 31.1 eV). During the one-hour interval, core electron and ion densities showed similar temporal variation with $n_{e}\simeq n_{p}$ \citep{Halekas_2022,Bercic_2020}. On the open magnetic field lines electron and proton temperatures were very close, while on closed magnetic field lines, the electron temperature increased notably, resulting in a higher $T_{e}/T_{p}$ ratio (1.2 to 1.5).

%

      \begin{figure}
   \centering
   \includegraphics[width=9 cm]{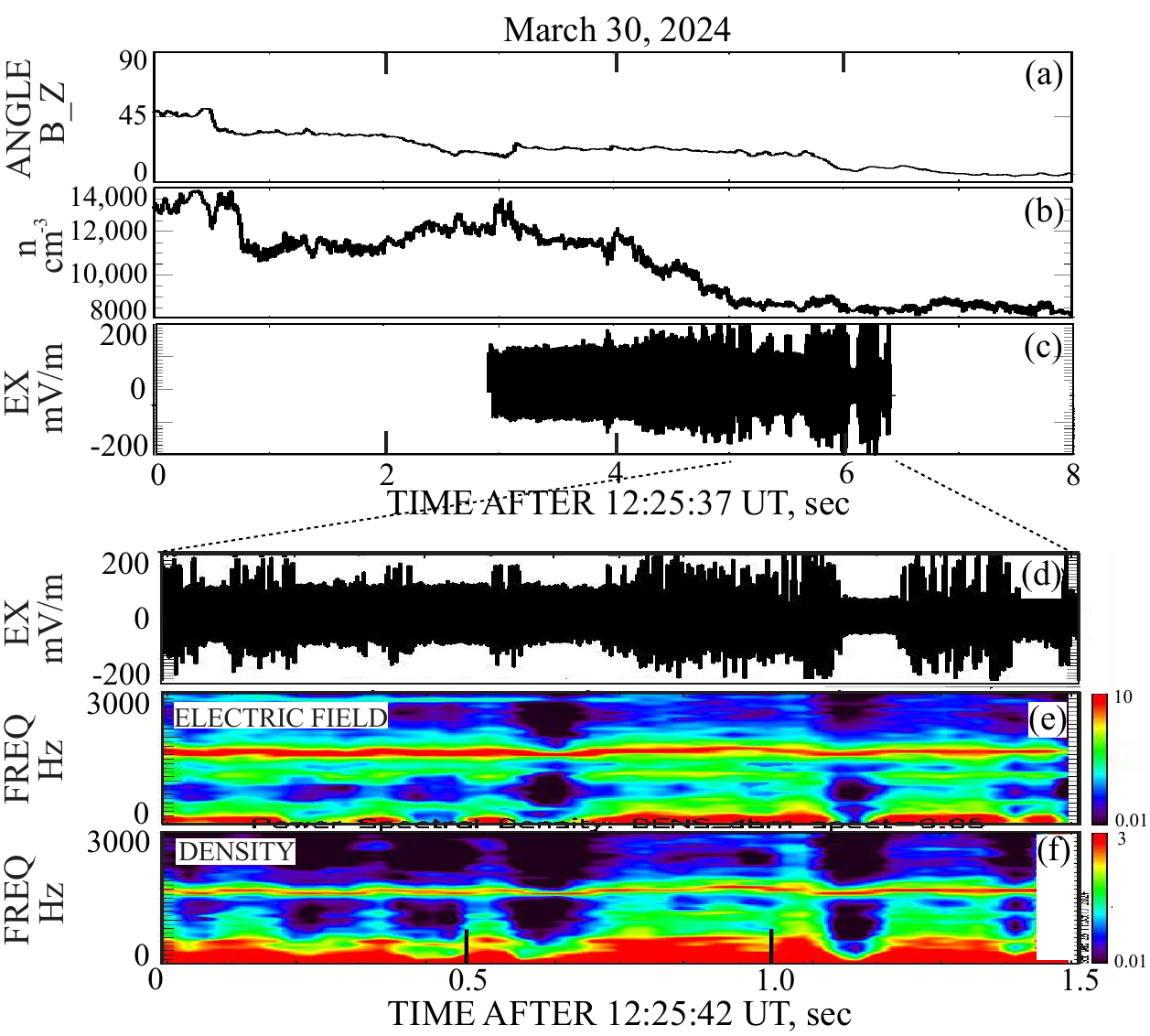}
      \caption{ An eight-second interval during which the magnetic field direction showed the spacecraft passage from a closed field-line-like geometry to an open geometry - (a). (b) - the density decreased synchronously with the changing magnetic field direction and there was a 3.5-second burst of high-rate electric field measurements shown in (c). (d) - one component of the electric field during this burst while the electric field spectrum from zero to 3000 Hz is displayed in (e) and the plasma density fluctuations from zero to 3000 Hz are illustrated in (f). The variations of the spectra on time scales shorter than 100 milliseconds suggest that the spacecraft crossed the intermingled open and closed magnetic field lines of a complex magnetic field geometry {that may have been} near the radial origin of the heliospheric current sheet. The similarity of the electric field and density spectra again shows that all the observed waves were electrostatic. 
              }
         \label{Fig3}
   \end{figure}
It is important to note that the only known mechanism for creation of the strahl is through interactions in the collisional plasma at the solar surface (see Appendix A) such that the identification of field lines as closed is unambiguous when bidirectional strahl flow is observed.  It is also important to note that such closed field lines were observed near 13 solar radii during the event of interest while a common belief has been that they do not exist beyond ~5 solar radii (Appendix A).

The interrelation between open and closed magnetic field lines on small spatial scales is revealed in Figure \ref{Fig3}, during which the spacecraft traveled from a closed-like to an open-like magnetic field geometry in eight seconds because the magnetic field direction in Figure \ref{Fig3}a changed from about 45 degrees to nearly zero with respect to the radial direction. In Figure \ref{Fig3}b the plasma density is shown to decrease in correlation with the magnetic field change, showing again that the large plasma density was a property of the closed, not open, field lines. Figure \ref{Fig3} c shows one component of the electric field collected in a burst that lasted about 3.5 seconds and is expanded in Figure \ref{Fig3}d. During this burst the electric field changed amplitude and character abruptly on time scales shorter than 100 milliseconds, which is a spatial scale about 1000 times the core electron gyroradius. The electric field spectrum to three kHz is shown in Figure \ref{Fig3}e while the spectrum of the fluctuations of the plasma density is shown in Figure \ref{Fig3}f \citep{Mozer_2022}. Both spectra change on time scales less than 100 milliseconds, showing that the waves were electrostatic. This also suggests that the open and closed magnetic field lines {comprise a} turbulent rather than well-defined static {geometry}. 

Figure {4} provides the relationship between the proton plasma frequency and the electric field harmonics over a three-hour interval surrounding the reversal of the radial magnetic field. The harmonics were visible for the full three-hour interval and their correlation with the proton plasma frequency (the white curve) is seen by the three black curves at 0.1$f_{pp}$, 0.2$f_{pp}$, and 0.3$f_{pp}$, which align nicely with the three lowest harmonics of the electric field. This is the expected correlation for ion acoustic waves. It is noted that the electron gyrofrequency, $f_{ce}$, illustrated as the red curve in the figure, is roughly anticorrelated with $f_{pp}$, so the harmonics do not correlate at all with $f_{ce}$.

      \begin{figure}
   \centering
   \includegraphics[width=\linewidth]{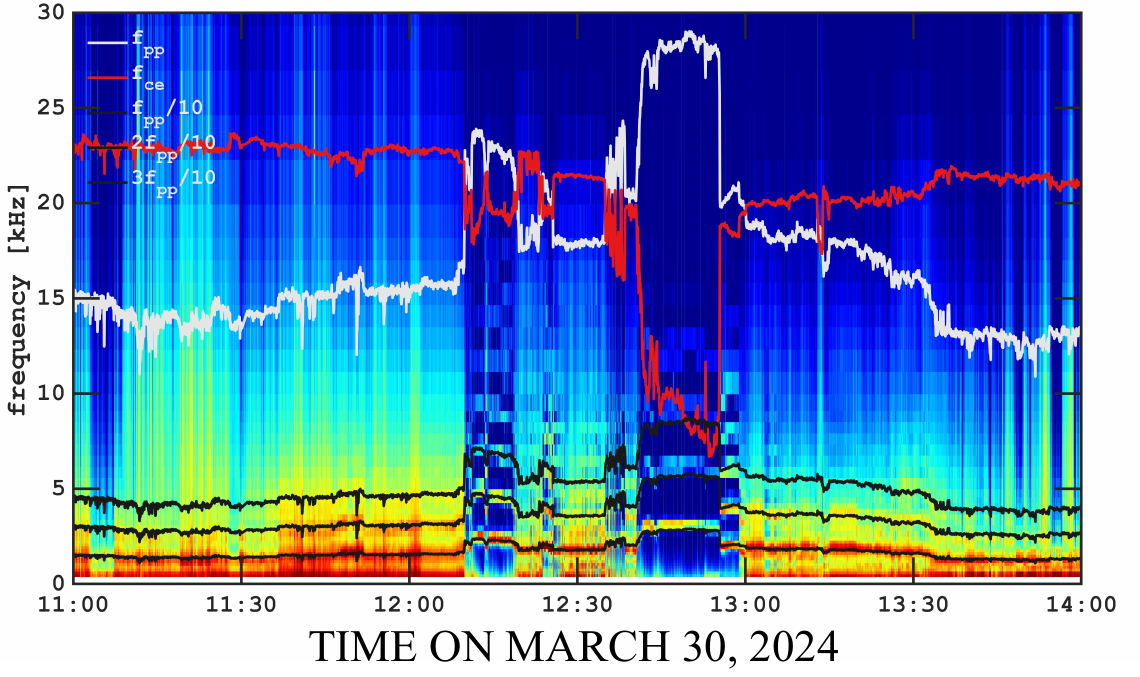}
      \caption{ The electric field spectra during the three-hour interval surrounding the reversal of the radial magnetic field.  The white curve gives the proton plasma frequency, $f_{pp}$, while the three black curves display 0.1$f_{pp}$, 0.2$f_{pp}$, and 0.3$f_{pp}$, respectively.  The red curve gives the gyrofrequency, $f_{ce}$. The correlation between the harmonic frequencies and the fractional proton plasma frequency is emphasized because this identifies the waves as ion acoustic waves. }
         \label{Fig4}
   \end{figure}

\section{Conclusions}
While crossing the heliospheric current sheet, the Parker Solar Probe (PSP) observed two distinct regions that contained widely different plasma densities, electric field spectra, fast changes of the magnetic field components, strahl electron flows and pitch angle distributions near 90 degrees.  These differences suggest that the spacecraft passed through both open and closed magnetic field lines in close proximity to the Sun - the heliocentric distance was 13 solar radii. The processing of fields and plasma perturbations in these two regimes revealed the following.

\begin{enumerate}
\item
The two regions intermingled on time scales of less than 100 milliseconds, creating a complex magnetic field geometry near the radial origin of the HCS. This rapid mixing highlights the dynamic nature of magnetic fields in solar plasma environments.
\item
The waves detected in both regions were identified as electrostatic. 
They comprised a broadband signal in regions with open magnetic field lines and well-structured harmonics of the main frequency in both open and closed field lines.
\item 
The observed harmonic frequencies were correlated with the plasma ion frequency ($f_{pp}$), with the main frequency at approximately  $0.1 f_{pp}$.
These characteristics indicate a strong relationship between wave phenomena and plasma properties.
\item
The waves with harmonics displayed field-aligned polarizations and were also observed as perturbations in plasma density. These attributes show that the observed electrostatic modes and their associated harmonics were ion acoustic waves.
\end{enumerate}

The absence of broadband electrostatic emissions in regions of closed magnetic field lines can be attributed to a lower ratio of hot to core electron density compared to open field regions. Additionally, a higher ratio of plasma frequency ($f_{pe}$) to cyclotron frequency ($f_{ce}$) in closed-field regions appears to suppress wave generation.


\backmatter

%
%
%

\bmhead{Acknowledgements}

This work was supported by NASA contract NNN06AA01C. OVA, KEC, LC, and RS  were supported by NASA contracts 80NSSC22K0522, 80NSSC22K0433, 80NNSC19K0848, 80NSSC21K1770, and NASA’s Living with a Star (LWS) program (contract 80NSSC20K0218). The authors acknowledge the extraordinary contributions of the Parker Solar Probe spacecraft engineering team at the Applied Physics Laboratory at Johns Hopkins University. The FIELDS experiment on the Parker Solar Probe was designed and developed under NASA contract NNN06AA01C. Our sincere thanks to J.W. Bonnell, M. Moncuquet, T. Quinn, M. Pulupa, and P. Harvey for providing data analysis material and for managing the spacecraft commanding and data processing, which have become heavy loads thanks to the complexity of the instruments and the orbit. 

%
%
%

%
%
%
%

\begin{appendices}

\section{Origin of the strahl and the radial extent of coronal loops}\label{secB}
\begin{figure}[hb]
   \centering
   \includegraphics[width=9cm]{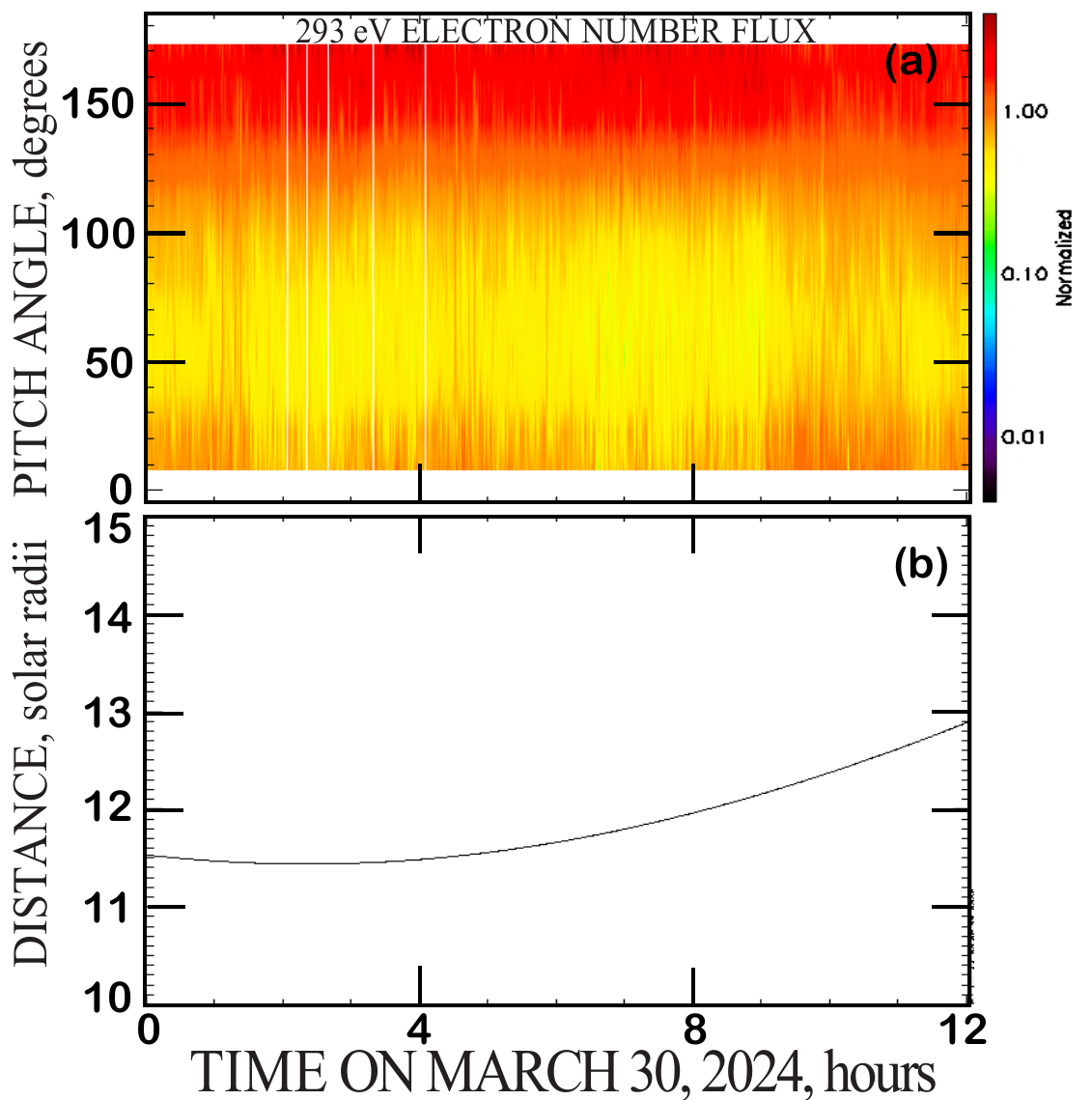}
      \caption{Panel (a) gives the 293 eV electron pitch angle distribution that peaks near both $0^\circ$ and $180^\circ$, which signify the presence of strahl and therefore, of closed field lines in coronal loops, for much of the 12-hour interval between 11.5 and 13 solar radii shown in panel (b).}
         \label{fig:B}
\end{figure}

Strahl, a beam of energetic electrons that propagate from the sun along Parker-spiral magnetic field lines, originates in the collisional photosphere \citep{boldyrev2019kinetic}.  Early modeling suggested that the coronal loops containing such strahl opened up by 2.5 solar radii \citep{schatten1969model,altschuler1969magnetic}, such that, at smaller distances, the closed field lines contained strahl flowing in both directions from the two ends of the field line and, beyond this distance, the field lines are open with only one end connected to the nearby solar surface, so they contained only outflowing strahl beams.  From such strahl observations one could therefore determine if the magnetic field line of interest was open or closed.
These early estimates of the maximum altitude of coronal loops that contained closed magnetic field lines, were extended by involving enhanced dynamics and complexities, \citep{panasenco2019large,reville2020tearing} to show that such closed field line regions can extend out to 10-12 solar radii.
For 12 hours on March 30, 2024 the Parker Solar Probe produced the electron pitch angle distribution at 293 eV shown in Figure~\ref{fig:B} panel (a).  That there were enhanced fluxes near both $0^\circ$ and $180^\circ$ shows that there was bi-directional strahl flow at radial distances to at least 13 solar radii, such that there must have been closed magnetic field lines and coronal loops to at least such altitudes.  This result both validates the reported observations of intermingled closed and open field lines at 13 solar radii and it invalidates the earlier idea that the tops of coronal loops were at altitudes below 5 solar radii. As an alternative explanation, one might imagine that bipolar strahl could appear on open field lines due to some type of reconnection with field lines carrying the strahl.  This possibility is refuted by the existence of bipolar strahl for 12 hours,  during which there was little or no reconnection.

\end{appendices}


\bibliography{bibliography}

\end{document}